\documentclass[tightenlines,aps,prc,showpacs,amsmath,showkeys,
superscriptaddress]{revtex4}
\usepackage[dvips]{graphicx}

\begin{document}

\title{\bf $\bar K$ nuclear bound states in a dynamical model}

\author{J.~Mare\v{s}}
\email{mares@ujf.cas.cz}
\affiliation{Nuclear Physics Institute, 25068 \v{R}e\v{z}, Czech Republic}

\author{E.~Friedman}
\email{elifried@vms.huji.ac.il}
\affiliation{Racah Institute of Physics, The Hebrew University,
Jerusalem 91904, Israel}

\author{A.~Gal}
\email{avragal@vms.huji.ac.il}
\affiliation{Racah Institute of Physics, The Hebrew University,
Jerusalem 91904, Israel}

\date{\today}

\begin{abstract}

A comprehensive data base of $K^-$-atom level shifts and widths
is re-analyzed in order to study the density dependence of the
$\bar K$-nuclear optical potential. Significant departure from 
a $t_{\rm eff}\rho$ form is found only for $\rho (r)/\rho_0~\lesssim~0.2$ 
and extrapolation to nuclear-matter density $\rho_0$ yields an attractive 
potential, about 170~MeV deep. Partial restoration of chiral symmetry 
compatible with pionic atoms and low-energy pion-nuclear data plays no 
role at the relevant low-density regime, but this effect is not ruled 
out at densities of order $\rho_0$ and beyond. 
$\bar K$-nuclear bound states are generated across the periodic table 
self consistently, using a relativistic mean-field model Lagrangian 
which couples the $\bar K$ to the scalar and vector meson fields 
mediating the nuclear interactions. 
The reduced phase space available for $\bar K$ absorption from these 
bound states is taken into account by adding an energy-dependent imaginary 
term which underlies the corresponding $\bar K$-nuclear level widths, with 
a strength required by fits to the atomic data. Substantial polarization 
of the core nucleus is found for light nuclei, and the binding energies 
and widths calculated in this dynamical model differ appreciably from 
those calculated for a static nucleus. A wide range of binding energies 
is spanned by varying the $\bar K$ couplings to the meson fields.
Our calculations provide a lower limit of $\Gamma_{\bar K} = 50 \pm 10$ 
MeV on the width of nuclear bound states for $\bar K$ binding energy in 
the range $B_{\bar K} \sim 100 - 200$ MeV. 
Comments are made on the interpretation of the FINUDA experiment at 
DA$\Phi$NE which claimed evidence for deeply bound $K^- pp$ states in 
light nuclei. 

\end{abstract}



\maketitle
\newpage

\section{Introduction}
\label{sec:int}

The existence of a $\bar K N$ unstable bound state, the $I=0$ 
$\Lambda(1405)$ $\pi \Sigma$ resonance, about 27 MeV below the $K^-p$ 
threshold was first discussed by Dalitz et al. \cite{DWR67} and shown 
to be compatible with strong coupled-channel vector-meson exchange 
$s$-wave interactions in the $I=0$ $\bar K N~-~\pi \Sigma$ system. 
This strongly attractive $I=0$ $\bar K N$ interaction, plus the moderately 
attractive $I=1$ $\bar K N$ interaction \cite{Mar76}, suggest that the 
$\bar K$-{\it nuclear} interaction is also strongly attractive. 
Indeed, the Born approximation for the $\bar K$-nucleus optical potential 
constructed from the leading-order Tomozawa-Weinberg (TW) vector coupling 
term of the  chiral effective Lagrangian \cite{WRW97}, 
\begin{equation} 
\label{eq:chiral} 
V_{\rm opt}^{\bar K}~ =~ - ~\frac{3}{8f_{\pi}^2}~\rho 
\end{equation} 
where $f_{\pi} \sim 93$ MeV is the pseudoscalar meson decay constant, 
yields sizable attraction $V_{\rm opt}^{\bar K} \sim -55$ MeV 
for a central nuclear density $\rho \sim \rho _0 = 0.16$ fm$^{-3}$. 
Iterating the TW term plus next-to-leading-order terms, 
within an {\it in-medium} coupled-channel approach constrained 
by the $\bar K N - \pi \Sigma - \pi \Lambda$ data near the 
$\bar K N$ threshold, roughly doubles this $\bar K$-nucleus attraction 
to about $-110$ MeV at $\rho _0$. It is found in these calculations 
(e.g. \cite{WKW96}) that the $\Lambda(1405)$ quickly dissolves in the 
nuclear medium at finite densities, well below $\rho _0$, so that 
the repulsive free-space scattering length $a_{K^-p}$ becomes 
{\it attractive}, and together with the weakly density dependent 
attractive $a_{K^-n}$ it yields a strongly attractive density-dependent 
$`t\rho'$ $\bar K$-nucleus optical potential: 
\begin{equation} 
\label{eq:trho} 
V_{\rm opt}^{\bar K}(r)~ = ~-{\frac{2\pi}{\mu_{KN}}}b_0(\rho)\rho(r)~, 
\end{equation} 
where $\mu_{KN}$ is the $\bar K N$ reduced mass and the in-medium 
isoscalar $\bar K N$ scattering length $b_0(\rho)$  
(positive for attraction in the sign convention common in this field) 
is defined by
\begin{equation} 
\label{eq:b0}
b_0(\rho)={\frac{1}{2}}(a_{K^-p}(\rho)+a_{K^-n}(\rho))~,
~~~~b_0(\rho_0) \sim 0.9~{\rm fm}~. 
\end{equation}
However, when $V_{\rm opt}^{\bar K}$ is calculated 
{\it self consistently}, namely by including $V_{\rm opt}^{\bar K}$ in the 
in-medium propagator used in the Lippmann-Schwinger equation determining 
$V_{\rm opt}^{\bar K}$ as a $\bar K$-nucleus T matrix, the resultant 
$\bar K$-nucleus potential is only moderately attractive, 
with depth between 40 and 60~MeV \cite{SKE00,ROs00,CFG01}. We call such 
$\bar K$-nucleus potentials `shallow'. In contrast, strongly attractive 
$\bar K$-nucleus potentials, of depth between 150 and 200 MeV, appear to
provide the best fit to the comprehensive $K^-$ atomic data which consist
of level shifts and widths across the periodic table
\cite{FGB93,FGB94,BFG97}. We call such $\bar K$-nucleus potentials 
`deep', noting that the fits provided by such `deep' potentials to the $K^-$ 
atomic data are by far superior to those provided by the `shallow' potentials 
\cite{CFG01}. 

Another approach to the construction of $V_{\rm opt}^{\bar K}$ in dense 
nuclear matter has been the relativistic-mean-field (RMF) approach 
where in addition to an attractive Lorentz-vector mean potential, 
similar in its origin to Eq. (\ref{eq:chiral}), an attractive Lorentz-scalar 
mean potential associated with the 
poorly known sigma term is included \cite{SGM94,SMi96,BRh96,FGM99}. This 
approach too yields a strongly attractive $V_{\rm opt}^{\bar K}$, the 
precise extent of attraction depending on the phenomenological constraints 
imposed on the RMF calculation. Assuming dominance of the Lorentz-vector 
mean potential proportional to $f_\pi^{-2}$, one could test the implications 
of partial restoration of chiral symmetry, in the sense originally proposed 
for pionic atoms \cite{Wei01}, by replacing in Eq. (\ref{eq:chiral}) $f_\pi$ 
by a density dependent decay constant $f_\pi^*$ according to 
\begin{equation} 
\label{eq:fpi2} 
f_\pi ^{*2} = f_\pi ^2 - \frac{\sigma }{m_\pi ^2} \rho ~~, 
\end{equation}
where $\sigma$ is the pion-nucleon sigma term \cite{GLS91}. 
This implies a specific density dependence of the optical potential 
(\ref{eq:trho}): \begin{equation}
\label{equ:WDD}
b_0(\rho)~=~\frac{b_0(0)}{1~-~0.046 \sigma \rho}
\end{equation}
where the $\pi N$ $\sigma$ term is given in MeV and $\rho$
is the nuclear density in fm$^{-3}$. This chirally motivated density 
dependence has proved successful in pionic atoms 
\cite{Fri02,Fri02a,KKW03,FGa03,FGa04,SFG04} and in low energy pion 
scattering \cite{FBB04,FBB05} where its effect can be separated from 
similar effects due to the threshold energy dependence of the underlying 
$\pi N$ amplitudes \cite{FGa04,FBB04}. 

The $\bar K$-nuclear interaction is also
strongly absorptive, which arises from the available one-nucleon
absorption reactions $\bar K N \rightarrow \pi Y$ with approximately
100 and 180 MeV energy release for the final hyperons ($Y$)
$\Sigma$ and $\Lambda$, respectively. Among the known hadron-nucleus 
interactions at low energies, only the $\bar N$-nucleus interaction 
is as strongly attractive and absorptive as the $\bar K$-nucleus 
interaction \cite{BFG95}. 

Recent experimental reports on candidates for $\bar K$-nuclear deeply
bound states in the range of binding energy 
$B_{\bar K} \sim 100 - 200$~MeV \cite{SBF04,SBF05,KHA05,ABB05} 
again highlighted the question of how attractive the $\bar K$-nucleus
interaction is below the $\bar K N$ threshold. Obviously, the `shallow'-type 
potentials cannot generate deeply bound nuclear states
in the energy range $B_{\bar K} \sim 100 - 200$ MeV, whereas the `deep' 
potentials might do. 
The distinction between `deep' and `shallow' $\bar K$-nucleus potentials
becomes somewhat fuzzy within a {\it dynamical} approach which allows
for polarization of the nucleus by the strong $\bar K$-nucleus interaction.
The depth of the $\bar K$-nucleus potential becomes then state dependent,
thus depending on the binding energy $B_{\bar K}$.
Indeed, strong polarization of the nucleus by the $\bar K$ interaction
was established in our recent RMF 
calculations for the light nuclei $^{12}$C and $^{16}$O \cite{MFG05}.
Consequences of such polarization are a central topic of the present work.

In Section 2 we re-analyze the existing $K^-$ atomic data in order to
gain information on the density dependence of the $\bar K$-nuclear
optical potential in the nuclear surface and to extrapolate smoothly
to nuclear matter density. This procedure yields a `deep'
attractive potential of about
170 MeV  at $\rho_0$. We also test whether starting from
a `shallow'-type potential \cite{SKE00,ROs00,CFG01},
and superimposing on it a density dependence of the form Eq. (\ref{equ:WDD}), 
could improve the fit to the $K^-$-atom data as it appears to do in 
pionic atoms and in low-energy pion-nuclear scattering 
\cite{Fri02,Fri02a,KKW03,FGa03,FGa04,SFG04,FBB04,FBB05}.
For $K^-$ atoms the answer is in the negative, 
perhaps because the nonperturbative aspect of the $\Lambda(1405)$ dominance 
at the low density regime encountered in kaonic atoms invalidates the 
chiral perturbation counting rules \cite{Wei05}. This
does not rule out a greater role for partial chiral restoration 
of a similar sort at higher densities, and for deeply 
bound $\bar K$-{\it nuclear} states, or for kaon condensation as discussed 
first by Brown et al. \cite{BLR94}. 

The case for $\bar K$-nucleus bound states with $B_{\bar K} \sim 100$ MeV,
generated by a `deep' potential, was made by Akaishi and Yamazaki
\cite{AYa02} who fitted the underlying $t$ matrix to a {\it partial} set of
the available low-energy $\bar K N$ data. For the most recent publication
that elaborates on their calculations for very light nuclear cores,
see Ref. \cite{ADY05}. In these few-body calculations, the strongly
attractive $\bar K$-nucleus interaction caused a substantial polarization,
or rearrangement of the nuclear core. In our {\it dynamical} RMF
calculations \cite{MFG05}, the $\bar K$ and the nucleons interact
through the exchange of scalar ($\sigma$) and vector ($\omega$) meson
fields which are treated in the mean-field approximation. The presence
of $\bar K$ induces additional source terms in the equations of motion
for the meson fields to which the $\bar K$ couples. This affects the
scalar and vector potentials which enter the Dirac equation for nucleons
and leads to rearrangement of the nuclear core.
It also leads to a deeper $\bar K$-nuclear potential with respect to
{\it static} calculations which assume a rigid nuclear core
(an assumption that is justified for $K^-$ and $\bar p$ atoms
\cite{MFG05,FGM05}). By successively allowing the $\bar K$ to polarize
the nucleons, and the polarized nucleus to enhance the $\bar K$-nuclear
interaction, this dynamical calculation is made self consistent.
Similar $\bar N$-nucleus RMF calculations have been reported by the 
Frankfurt group \cite{BMS02,MSB05}. 

In the present paper we extend the results of Ref. \cite{MFG05}
in order to study the dynamical aspects of the $\bar K$-nucleus coupling
across the periodic table, and in particular to study the widths expected
for deeply bound states over a wide range of binding energies. This is
done by scanning on the coupling constants of the $\bar K$ - meson fields.
In Section 3 we outline the $\bar K$-nucleus RMF methodology used in
this work and discuss its extension to describe absorptive interactions
by which the $\bar K$-nuclear bound states acquire a width. The calculation
of the width is also done dynamically, since its evaluation requires the
nonstatic nuclear density and the binding energy which determines the
reduced phase space available to the absorption reactions, and both of these
quantities keep changing from one step of the iterative solution of the
equations of motion to the next. The imaginary part of the potential
affects also nonnegligibly the calculated binding energy. This important
aspect of the $\bar K$-nucleus dynamics is missing in the very recent RMF
calculation by Zhong et al. \cite{ZPL05}. In Section 4 we show and
discuss the results of calculations across the periodic table.
Special attention is paid to $^{12}$C and other
light nuclear targets for which
a recent FINUDA experiment at DA$\Phi$NE, Frascati, claimed evidence for
deeply bound $K^- pp$ states \cite{ABB05}. We find large dynamical effects
for light nuclei such as $^{12}$C and we discuss these in detail.
Section 5 summarizes the present work with conclusions and outlook.

\section{Kaonic atoms}
\label{sec:katoms}

\subsection{Motivation and background}

As explained in the Introduction, the motivation for re-analysis of a
comprehensive set of kaonic atoms data is two fold. First is the question 
of `deep' {\it vs.} `shallow' real $\bar K$-nucleus potential,
in light of recent possible experimental evidence for the existence of deeply
bound kaonic states whose binding energies exceed the depth of the shallow
type of potential obtained from fits to kaonic atoms data. Such shallow
potentials are obtained when fits to the data are made
with a simple `$t \rho$' approach, as described below,
and also when self-consistent approaches are used \cite{SKE00,ROs00,CFG01}.
However, if the deep variety of potential is confirmed, then the
dependence of the $\bar K N$ interaction on the nuclear density becomes
of prime concern. This density dependence is
the second point which motivated the re-analysis of kaonic atoms data
although there have not been any new experimental results on strong
interaction effects in kaonic atoms since the early 1990's.

The method adopted here is rather similar  to our earlier
work \cite{FGB93,FGB94,BFG97}, namely, performing global fits to a large
set of data which covers the whole of the periodic table, using an
optical potential within the Klein-Gordon equation,
\begin{equation}
\label{eq:KG1}
\left[\Delta - 2{\mu}(B+V_{\rm opt}+V_c) + (V_c+B)^2 \right]{\psi} = 0~~ ~~
(\hbar = c = 1),
\end{equation}
where $V_c$ denotes the static Coulomb potential for the $\bar K$ due to
the finite charge distribution of the nucleus, including the first-order
$\alpha (Z\alpha)$ vacuum-polarization potential,
$\mu$ is the $\bar K$-nucleus reduced mass and
$B=B_{\bar K}+{\rm i}\Gamma_{\bar K}/2$,
with $B_{\bar K}$ and $\Gamma_{\bar K}$ standing for the binding energy
and width, respectively.
The interaction of $\bar K$ with the nucleus is described here in terms
of an optical potential $V_{\rm opt}$ which in the simplest
`$t \rho$' form is given by
\begin{equation}
\label{equ:potl}
2\mu V_{{\rm opt}}(r) = -4\pi(1+\frac{\mu}{m_N}
\frac{A-1}{A})b_0(\rho_n(r)+\rho_p(r))~~,
\end{equation}
where $\rho_n$ and $\rho_p$ are the neutron and proton density
distributions normalized to the number of neutrons $N$ and number
of protons $Z$, respectively, $A=N+Z$, and $m_N$ is the mass of the
nucleon. In the impulse approximation `$t \rho$' approach the parameter
$b_0$ is equal to the $\bar K$-nucleon isoscalar scattering length 
under the sign convention common in this field, otherwise this parameter
may be regarded as `effective' and its value is obtained from fits to 
the data. For $A~ >>~ 1$, Eq. (\ref{equ:potl}) reduces to 
Eq. (\ref{eq:trho}) discussed above in the context of nuclear matter. 
An isovector term is not included in Eq. (\ref{equ:potl}) 
as earlier analyses showed 
\cite{BFG97} that such a term is not required by fits to kaonic atoms data.
Compared to the previous analyses we have now used a modified set of
parameters for the nuclear densities.
The density distribution of the protons is usually considered known as
it is obtained from the nuclear charge distribution by
unfolding the finite size of the charge of the proton.
The neutron distributions are, however, generally not known to sufficient
accuracy.
Experience with pionic atoms \cite{FGa03} and with antiprotonic atoms
\cite{FGM05} showed that
the feature of neutron density distributions which is most relevant
in determining strong interaction effects
in exotic atoms is the radial extent, as represented for
example by $r_n$, the neutron density rms radius. Other features
such as the detailed shape of the distribution have only minor effect.
This is observed also for kaonic atoms where the
dependence of the quality of fits on the densities
of the neutrons is less than what is observed for pionic and antiprotonic
atoms. The neutron densities used in the present work were of the 
two-parameter Fermi shape and of the `skin' variety, as detailed in 
Ref.~\cite{FGM05}. The values of $r_n$ were those that yielded the best 
fit to pionic and antiprotonic atoms data.

With the potential Eq. (\ref{equ:potl}) and varying the
complex parameter $b_0$ we
obtain the same results as before \cite{BFG97} when the data span a range
of nuclei from $^7$Li to $^{238}$U. The value of $\chi ^2$ is marginally
larger than the corresponding value found in \cite{BFG97} because of the
slightly different nuclear densities used in the two calculations.
The depth of the real potential for a typical medium-weight to heavy
nucleus is about 80 MeV. Recall that the corresponding value for the 
shallow-type potentials is close to 55 MeV \cite{CFG01}.

\subsection{Density dependence}

The possibility of departure of the antikaon-nucleus potential from the
simple `$t \rho$' form had been studied over a decade
ago \cite{FGB93,FGB94,BFG97}
and significant improvement in the fit to the data had been achieved.
The empirical parameter $b_0$ was made density dependent (DD) by 
replacing $b_0$ in Eq. (\ref{equ:potl}) by 
\begin{equation}
\label{equ:DD1}
b_0~\rightarrow b_0~+~B_0~[\frac{\rho(r)}{\rho _0}]^\alpha~
\end{equation}
where $\rho _0~ =~ 0.16~{\rm fm}^{-3}$. 
In this way it was possible to respect the `low density limit'
(when $\alpha >~0$) by keeping
$b_0$ fixed at its free $\bar K N$ value and varying the parameters $B_0$
and $\alpha$. In view of the importance of the modification
of the $\bar K N$ interaction in nuclear matter we have addressed
this question again in the present work in a different way.

We {\it loosely} define in coordinate space an `internal' region and
an `external' region by using the multiplicative functions $F(r)$ in
the former and $[1-F(r)]$ in the latter, where $F(r)$ is defined as
\begin{equation}
\label{equ:F}
F(r)~=~\frac{1}{e^x +1}
\end{equation}
with $x~=~(r-R_x)/a_x$. Then clearly $F(r)~\rightarrow~1$ when
$r~<~R_x~-~3~a_x$, which is the internal region. Likewise
$[1~-~F(r)]~\rightarrow~1$ in the external region.
Thus $R_x$ forms an approximate border between the internal and the external
regions, and {\it if} $R_x$ is close to the nuclear surface then
the two regions will correspond to the high density and low density
regions of nuclear matter, respectively.
We have therefore replaced the fixed $b_0$
in the $t\rho$ potential Eq. (\ref{equ:potl}) by
\begin{equation}
\label{equ:DDF}
b_0~\rightarrow ~B_0~F(r)~+~b_0~[1~-~F(r)]~~.
\end{equation}
Here the parameter $b_0$ represents the large-$r$ interaction, and
it can be held fixed, e.g.
at the free $\bar K N$ value. The parameter $B_0$ represents
the interaction
inside the nucleus. As stated above, this division into regions
of high and low densities is meaningful
{\it provided} $R_x$ is close to the radius of the nucleus
and $a_x$ is of the order of 0.5 fm. To enable global fits to be made
over the whole of the periodic table we parameterized $R_x$ as
$R_x~=~R_{x0}~A^{1/3}~+~\delta _x$ and varied in the least-squares fit the
values of $B_0,~ R_{x0}$ and $\delta _x$ while gridding on values of $a_x$.
The  parameter Re$b_0$ was held fixed at its free $\bar K N$ value
but the results depend very little on its precise value.

\begin{table}
\caption{Parameter values from global fits to kaonic atoms data, for the 
models of Eqs. (\ref{equ:potl})-(\ref{equ:DDF}). Additional parameters for 
the fit using Eq. (\ref{equ:DDF}) are $R_{x0}=1.30\pm0.05$ fm, 
$\delta _x=0.8\pm0.3$ fm, $a_x$=$\underline {0.4}$ fm.
Underlined parameter values were held fixed in the fit process. $\chi ^2$
values are for 65 data points, $b_0$ and $B_0$ are in units of fm.}
\label{tab:katoms}
\begin{ruledtabular}
\begin{tabular}{lcccccc}
potential&$\chi ^2$&Re$b_0$&Im$b_0$&Re$B_0$&Im$B_0$&$\alpha$ \\ \hline
Eq. (\ref{equ:potl}) (`$t\rho$')&129.4&0.62$\pm$0.05&0.92$\pm$0.05&-&-&-  \\
Eq. (\ref{equ:DD1}) (`DD')&103.3&$\underline{-0.15}$&$\underline{0.62}$&
  1.62$\pm$0.07&$-$0.02$\pm$0.02&0.24$\pm$0.03 \\
Eq. (\ref{equ:DDF}) (`F')& 84.4&$\underline{-0.15}$&$\underline{0}$&
1.42$\pm$0.05&0.66$\pm$0.05& - \\
\end{tabular}
\end{ruledtabular}
\end{table}

Table \ref{tab:katoms} shows results of $\chi ^2$ fits to kaonic atoms
data over the whole of the periodic table using the three potentials
described above. For the third option, Eq. (\ref{equ:DDF}),
the values of the geometrical parameters for $a_x=0.4$ fm 
are found to be
$R_{x0}=1.30\pm0.05$ fm, $\delta _x=0.8\pm0.3$ fm, implying that the
modification of the free $\bar K N$ interaction takes place at radii
somewhat outside of the nuclear `half-density' radius.
Changing the value of $a_x$ between 0.2 and 0.5~fm made no difference.
The ansatz Eq. (\ref{equ:DDF}) was applied to both real and imaginary parts
of $B_0$, otherwise the fit deteriorates.

\begin{figure}
\includegraphics[height=9cm,width=10cm]{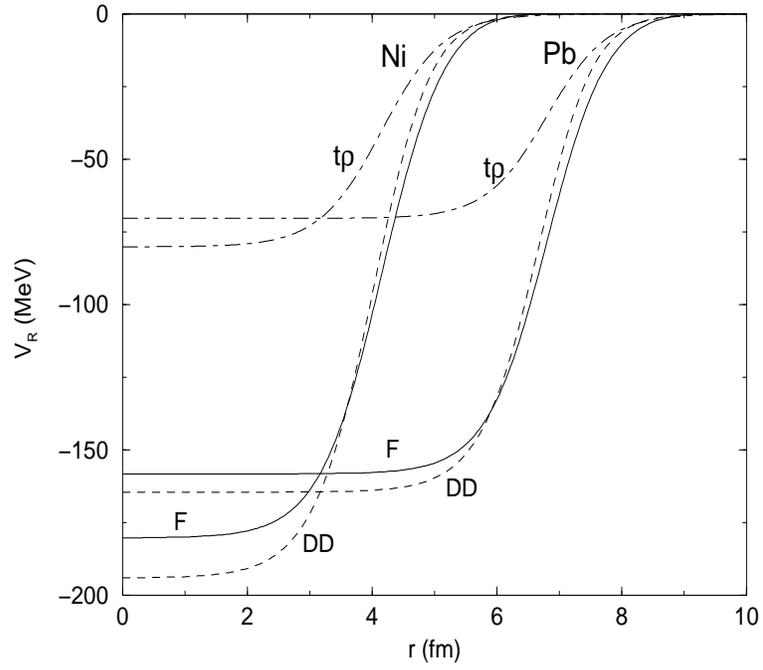}
\caption{Real part of the $\bar K$-nucleus potential for $^{58}$Ni and 
$^{208}$Pb, for the $t\rho$ model ($t\rho$, dash-dots), for the DD 
potential of Eq.~(\ref{equ:DD1}) (DD, dashed) and for the potential of 
Eq.~(\ref{equ:DDF}) (F, solid curves).}
\label{fig:VR}
\end{figure}

\begin{figure}
\includegraphics[height=9cm,width=10cm]{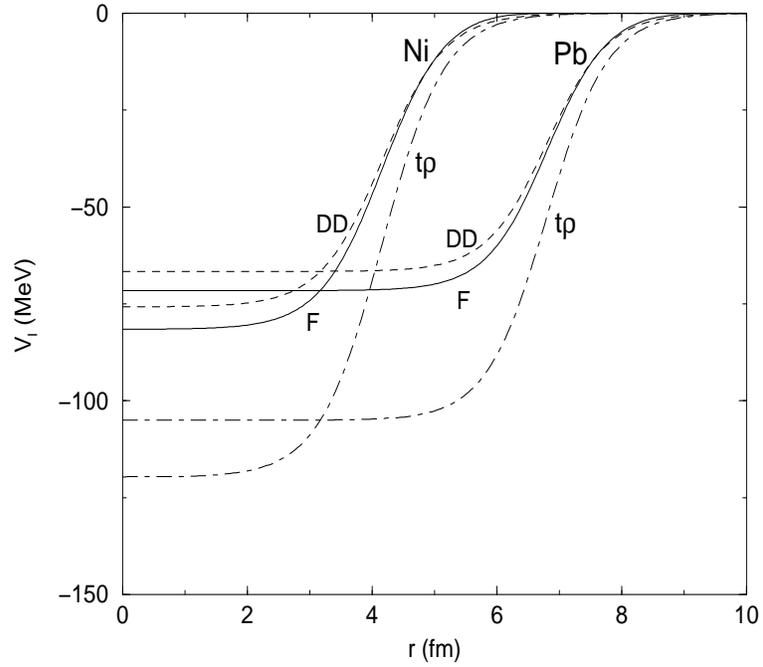}
\caption{Same as Fig.~\ref{fig:VR} but for the imaginary part of the 
$\bar K$-nucleus potential.}
\label{fig:VI}
\end{figure}

Figure \ref{fig:VR} shows the real part of the $\bar K$-nucleus
potential for $^{58}$Ni
and $^{208}$Pb for the $t\rho$ model and for the two density-dependent
potentials described above. The difference between the relatively shallow 
$t\rho$ potential and the deep potentials is clearly seen. 
Figure \ref{fig:VI} shows similar results for the imaginary part of 
the potential. Although the two density-dependent potentials have very 
different parameterizations the resulting potentials are quite similar.

\begin{figure}
\includegraphics[height=8cm,width=10cm]{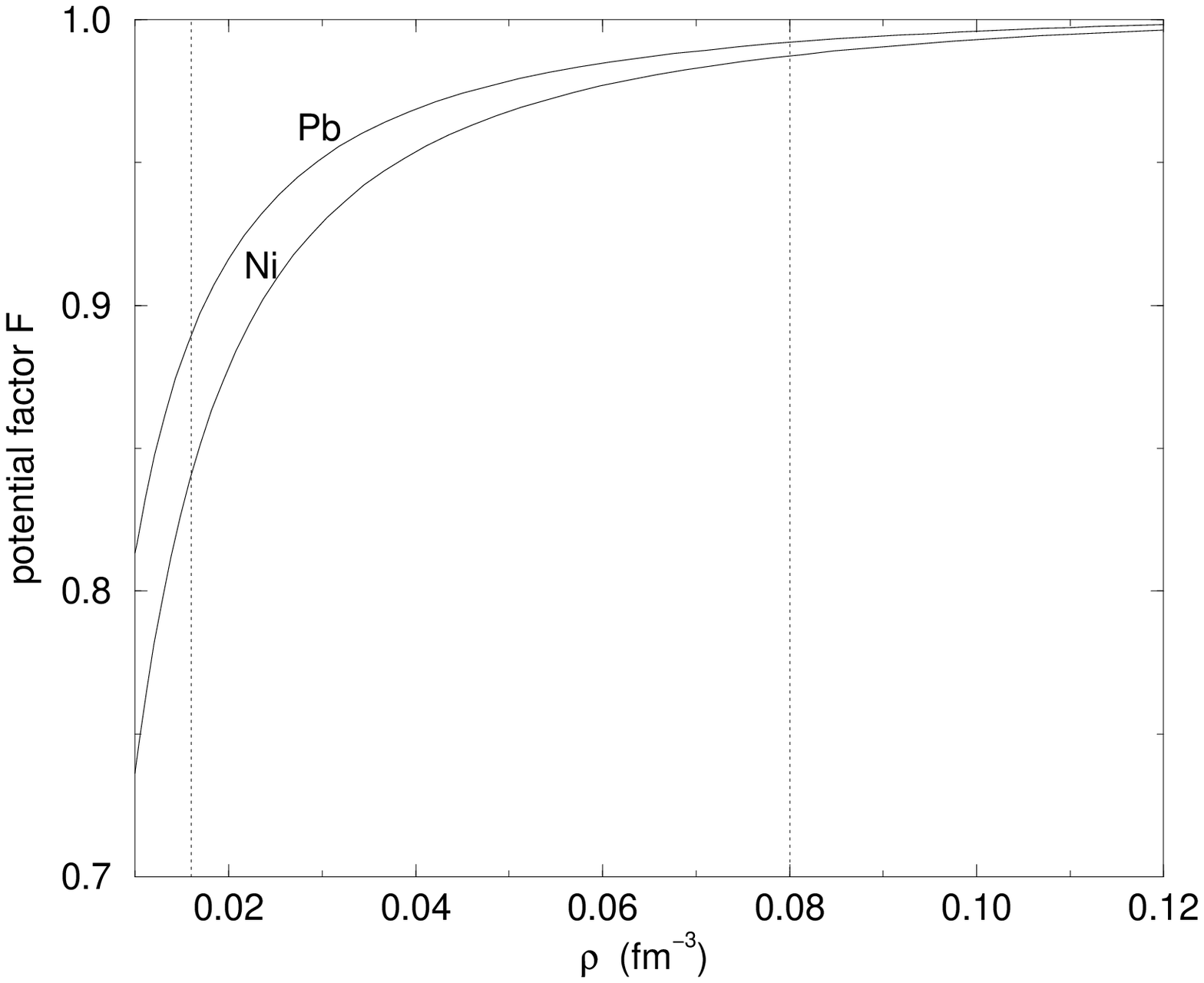}
\caption{The modifying factor $F$, Eq. (\ref{equ:F}), as a functional of 
$\rho$, see text. The vertical dotted lines indicate 10\% and 50\% of 
nuclear matter density $\rho_0 = 0.16$ fm$^{-3}$.}
\label{fig:F}
\end{figure}

In Figure \ref{fig:F}
are plotted the $F(r)$ functions for $^{58}$Ni and $^{208}$Pb, not
as functions of $r$ but rather as {\it functionals} of the local nuclear
density $\rho$, which is of prime interest in the present
work. We note that for these two examples, representing
medium-weight and heavy nuclei, the two curves are very close to each
other. The two vertical dotted lines represent 10\% and 50\% of a nuclear
matter density of 0.16 fm$^{-3}$. It is evident that the potential departs
from the $\rho (r)$ dependence for density  lower than
about 20\% of its central value.

Finally, we have tried to fit the $K^-$ atoms data with the  
density dependence given by Eq. (\ref{equ:WDD}) as a possible signature 
of partial restoration of chiral symmetry in dense 
nuclear medium \cite{Wei01}. 
However, no improvement on the shallow $t\rho$ potential for $K^-$ atoms 
was achieved when starting from the self-consistent amplitudes
of Ramos and Oset \cite{ROs00}
using for $\sigma$ values around 50 MeV \cite{GLS91}.
That is in line with Fig. \ref{fig:F} where it
is seen empirically that the density dependence is effective
only at fairly low densities, considerably lower than the ones reached 
in pionic atoms ($\sim 0.5~\rho_0$). 

\section{$\bar K$ Nuclear RMF Methodology}
\label{sec:meth}

The theoretical framework adopted in this work for calculating $\bar K$ 
nuclear bound states is the relativistic mean field (RMF) model for 
a system of nucleons and one $\bar K$ meson \cite{MFG05}. In this model, 
the interactions between hadrons are mediated by the exchange of scalar 
and vector boson fields which are treated in the mean-field approximation.
In the calculations reported in Ref. \cite{MFG05} and extended here, 
we employed the standard RMF Lagrangian ${\cal L}_N$ for the description 
of the nucleon sector, with the linear parameterization L-HS of Horowitz 
and Serot \cite{HSe81}, as well as with several nonlinear 
parameterizations, but mostly NL-SH due to Sharma et al. \cite{SNR93}. 
These parameterizations give quite different values for the nuclear
compressibility and, therefore, are expected to provide different
rearrangement of the nuclear core due to the presence of $\bar K$.
The (anti)kaon interaction with the nuclear medium is
incorporated by adding to ${\cal L}_N$ the Lagrangian density
${\cal L}_K$ \cite{SGM94,SMi96}:
\begin{equation}
\label{eq:Lk}
{\cal L}_{K} = {\cal D}_{\mu}^*{\bar K}{\cal D}^{\mu}K -
m^2_K {\bar K}K
- g_{\sigma K}m_K\sigma {\bar K}K\; .
\end{equation}
The covariant derivative
${\cal D_\mu}=\partial_\mu + ig_{\omega K}{\omega}_{\mu}$ describes
the coupling of the (anti)kaon to the vector meson $\omega$.
The vector field $\omega$ is then associated with a conserved current.
The coupling of the (anti)kaon to the isovector $\rho$ meson is here
excluded. This is a good approximation for $N=Z$ nuclei which holds for 
the majority of nuclei considered here, but not necessarily for the
heavier $^{208}$Pb nucleus. We note in this respect that
optical-potential fits as described in Sect. \ref{sec:katoms} for kaonic
atoms, including data for nuclei with excess neutrons up to $^{238}$U,
found no need to introduce isovector components.

Whereas extending the nuclear Lagrangian ${\cal L}_N$ by the Lagrangian
${\cal L}_K$ does not affect the original form of the corresponding Dirac
equation for nucleons, the presence of $\bar K$ induces additional source
terms in the equations of motion for the meson fields $\sigma$ and
$\omega_0$ to which the $\bar K$ couples:
\begin{eqnarray}
\left(-\Delta + m_{\sigma}^{2} \right) \sigma \; &=&
- g_{\sigma N}\rho_{S}
- g_{\sigma K} m_K {\bar K}K
+ \left( - g_{2}\, \sigma^{2} -
g_{3}\, \sigma^{3} \right) \;,
\label{eq:sigma}
\end{eqnarray}
\begin{eqnarray}
\left(-\Delta + m_{\omega}^{2} \right) \omega_0 &=&
+ g_{\omega N}\rho_{V}
- 2 g_{\omega K} (\omega_K+g_{\omega K}\omega_0) {\bar K}K \; ,
\label{eq:omega}
\end{eqnarray}
where $\omega_K$ is the $\bar K$ energy in the nuclear medium:
$$
\omega_K = \sqrt{m^2_K + g_{\sigma K} m_K \sigma + p^2_K}
- g_{\omega K}\omega_0\; ,
$$
and $\rho_{S}$ and $\rho_{V}$ denote the nuclear scalar and vector
densities, respectively. The additional source terms due to the $\bar K$
in Eqs. (\ref{eq:sigma})-(\ref{eq:omega})
affect the scalar and vector fields (potentials) which enter the
Dirac equation for nucleons. This leads to the rearrangement,
or polarization of the nuclear core in the presence of $\bar K$.

The equation of motion for $\bar K$ is derived from the
Lagrangian ${\cal L}_K$ using standard techniques. 
In order to preserve the connection to previous studies of kaonic
atoms, the corresponding Klein Gordon (KG) equation of motion for the
$\bar K$ is used in the form of Eq. (\ref{eq:KG1}) which is rewritten here
\begin{equation}
\label{eq:KG2}
\left[\Delta - 2{\mu}(B^{\rm s.p.}+V_{\rm opt}+V_c) + (V_c+B^{\rm s.p.})^2
\right]{\bar K} = 0~~ ~~ (\hbar = c = 1)~,
\end{equation}
where the superscript s.p. in
$B^{\rm s.p.}=B_{\bar K}^{\rm s.p.}+{\rm i}{\Gamma_{\bar K}}/2$
stands for the single-particle $\bar K$ binding energy which is equal
to the $\bar K$ separation energy only in the static calculation, as
elaborated below.
Moreover, in this particular form of the KG equation, it admits a
straightforward extention, by allowing $V_{\rm opt}$ to become {\it complex},
to consider dynamically the width $\Gamma_{\bar K}$ of $\bar K$ nuclear
bound states. This is a new and crucial element in our calculations.
The real part of the $\bar K$ optical potential $V_{\rm opt}$ in
Eq.(\ref{eq:KG2}) is then given by
\begin{equation}
\label{eq:VOP1}
{\rm Re}V_{\rm opt} = {\frac{m_K}{\mu}} [{\frac{1}{2}}S - 
(1 - \frac{B_{\bar K}^{\rm s.p.} + V_c}{m_K})V - {\frac{V^2}{2m_K}}]
\; ,
\end{equation}
where $S = g_{\sigma K}\sigma$ and $V = g_{\omega K}\omega_0$ are the
scalar and vector potentials due to the $\sigma$ and $\omega$ mean fields,
respectively. We note that Re$V_{\rm opt}$ is explicitly 
{\it state dependent} through the $(1 - [B_{\bar K}^{\rm s.p.} + V_c]/m_K)$ 
energy-dependent factor which multiplies the vector potential. 
This factor was disregarded in our Letter \cite{MFG05} where the state 
dependence of Re$V_{\rm opt}$ arose only implicitly through the dynamical 
density dependence of the mean-field potentials $S$ and $V$. 

The binding (separation) energy of $\bar K$, $B_{\bar K}$, in the combined
$\bar K$ nuclear system $^A_{\bar K}Z$ is given by
\begin{equation}
\label{eq:BK}
B_{\bar K} = B(^A_{\bar K}Z) - B(^AZ)~.
\end{equation} 
Here $B(^AZ)$ is the {\it total} binding energy of the nucleus $^AZ$. 
It is to be noted that the total binding energy $B(^A_{\bar K}Z)$ of the
$\bar K$ nucleus, besides including nucleon single-particle energies,
mean-field energies and center-of-mass energy, also includes the
$\bar K$ single-particle binding energy $B_{\bar K}^{\rm s.p.}$ from
Eq. (\ref{eq:KG2}) and the additional meson mean-field contributions
from the source terms due to (anti)kaons in the KG 
Eqs. (\ref{eq:sigma})-(\ref{eq:omega}) for the $\sigma$ and $\omega$ meson 
fields. Whereas $B_{\bar K} = B_{\bar K}^{\rm s.p.}$ for the static 
calculation, $B_{\bar K}^{\rm s.p.} > B_{\bar K}$ for the dynamical 
calculation, and this latter difference defines the nuclear rearrangement 
energy which is related to the polarization of the nuclear core by the 
$\bar K$. These statements are demonstrated in Section \ref{sec:res}. 

Since the traditional RMF approach does not address the imaginary part
of the potential, ${\rm Im}V_{\rm opt}$ was taken in a phenomenological
$t\rho$ form, where its depth was fitted to the $K^-$ atomic data
\cite{FGM99} and the nuclear density $\rho$ was calculated within the RMF
model. We emphasize that $\rho$ in the present calculations is no longer
a static nuclear density. Here it is a {\it dynamical} entity affected
by the $\bar K$, which is embedded in the nuclear medium and interacts
with the nucleons via boson fields. The resulting increased nuclear
density leads to increased widths, particularly for deeply bound states.
On the other hand, the phase space available for the decay products is
reduced for deeply bound states, which will act to decrease the calculated
width. Thus, suppression factors multiplying
${\rm Im}V_{\rm opt}$ were introduced from phase-space considerations,
taking into account the binding energy of the kaon for the initial decaying
state, and assuming two-body final-state kinematics for the decay products.
Two absorption channels were considered. The dominant one for absorption
at rest is due to pionic conversion modes on a single nucleon:
\begin{equation}
\label{eq:conv1}
{\bar K}N\rightarrow \pi\Sigma,\;\; \pi\Lambda \;\;\;\; (\sim 80\%)\;\;,
\end{equation}
with thresholds about 100 MeV and 180 MeV, respectively, below the
${\bar K} N$ total mass.
The corresponding density-independent suppression factor is given by
\begin{equation}
\label{eq:spf1}
f_1=\frac{M_{01}^3}{M_1^3}\sqrt{\frac{[M_1^2-(m_\pi +m_Y)^2]
[M_1^2-(m_Y-m_\pi)^2]}
{[M_{01}^2-(m_\pi +m_Y)^2][M_{01}^2-(m_Y-m_\pi)^2]}}~
\Theta (M_1-m_{\pi}-m_Y)  \;\;\; ,
\end{equation}
where $M_{01}=m_K+m_N,~M_1=M_{01}-B_{\bar K}$.
The second channel is due to non-pionic absorption modes on
two nucleons:
\begin{equation}
\label{eq:conv2}
{\bar K} N N \rightarrow Y N \;\;\;\; (\sim 20\%)\;\;,
\end{equation}
with thresholds about $m_{\pi}=140$ MeV lower than the single-nucleon
threshold. This second channel represents in our model all the
{\it multi-nucleon} absorption modes which are not resolved by experiment.
The corresponding suppression factor is given by
\begin{equation}
\label{eq:spf2}
f_2=\frac{M_{02}^3}{M_2^3}\sqrt{\frac{[M_2^2-(m_N +m_Y)^2]
[M_2^2-(m_Y-m_N)^2]}
{[M_{02}^2-(m_N +m_Y)^2][M_{02}^2-(m_Y-m_N)^2]}}~
\Theta (M_2-m_Y-m_N)  \;\;\; ,
\end{equation}
where $M_{02}=m_K+2m_N,~M_2=M_{02}-B_{\bar K}$.
The branching ratios (quoted above in parentheses) are known from
bubble-chamber experiments \cite{VVW77}.
Although multi-nucleon absorption modes are often modeled to have
a power-law ${\rho}^{\alpha}$ ($\alpha$ $>$ 1) density dependence,
a linear density dependence ($\alpha = 1$) may also arise in multi-step
reaction mechanisms \cite{Kol79}. Since our comprehensive $K^-$-atom
fits \cite{FGB93} are satisfied with $\alpha \sim 1$, we here
assume $\alpha = 1$ which means that $f_2$ too is independent of density.
We comment below on the effect of a possible density dependence of
$f_2$, reflecting perhaps a ${\rho}^2$ dependence of the non-pionic
decay mode Eq. (\ref{eq:conv2}) at high densities.

Since $\Sigma$ final states dominate both the pionic and non-pionic
channels \cite{VVW77}, the hyperon $Y$ was here taken as $Y=\Sigma$.
Allowing $\Lambda$ hyperons would foremost {\it add} conversion width
to $\bar K$ states bound in the region $B_{\bar K} \sim 100 - 180$ MeV.
For the combined suppression factor we assumed a mixture of 80\%
mesonic decay and 20\% nonmesonic decay \cite{VVW77}, i.e.
\begin{equation}
\label{eq:spf}
f=0.8~f_1~+~0.2~f_2 \; .
\end{equation}
In the calculations below, a residual value of $f=0.02$ was assumed
when both $f_1$ and $f_2$ vanish.

The coupled system of equations for nucleons and for the electromagnetic
vector field $A_0$, for the $\rho$ meson mean field, and for the mean fields 
$\sigma$ and $\omega_0$ Eqs. (\ref{eq:sigma})-(\ref{eq:omega}) above, 
as well as the KG Eq. (\ref{eq:KG2}) for $K^-$, were solved 
self consistently using an iterative procedure.
Obviously, the requirement of self-consistency is crucial for the proper
evaluation of the dynamical effects of the $\bar K$ on the nuclear
core and vice versa. We note that self consistency is not imposed
here on the final-state hadrons which only enter through their
{\it on-shell} masses used in the phase-space suppression factors
given above. For the main $\pi \Sigma$ decay channel it is likely
that the attraction provided by the pion within a dynamical calculation
\cite{ROs00} is largely cancelled by the nuclear repulsion
deduced phenomenologically for $\Sigma$ hyperons \cite{MFG95,NSA02,SNA04}.

\section{Results and Discussion}
\label{sec:res}

\subsection{Bulk properties of $K^-$ nuclear bound states}
\label{bulk}

In this subsection we show and discuss results of calculations
of `bulk' properties for $K^-$ nuclear bound states,
such as level widths, average nuclear densities and rms radii,
and single particle energies. The calculations cover a wide range
of binding energies, in order to establish correlations between some
of these properties and to study effects due to the nuclear polarization.
The empirical values $g^{(1)}_{\sigma K}$ and $g^{(1)}_{\omega K}$ for the
RMF $\bar K$ coupling constants, as found from a fit to kaonic atom data
(third and fourth lines in Table I of Ref. \cite{FGM99}), 
were used as a starting point for calculations.
A full dynamical calculation was then made for $K^-$ {\it nuclear} states
starting from the static $t\rho$ imaginary potential obtained from the
atomic fit with Im$b_0$ = 0.62~fm from Table \ref{tab:katoms},
while entering dynamically in the iteration cycles the resulting nuclear
density and the suppression factor $f$ as defined by Eq.(\ref{eq:spf}).  
This {\it dynamical} calculation, for the light nuclei $^{12}$C and $^{16}$O, 
led to a substantial increase of the calculated binding energy. For example, 
whereas the static calculation gave $B_{K^-} = 132$ MeV for the $1s$ state 
in $^{12}$C, the dynamical calculation gave $B_{K^-} = 172$ MeV for this same 
state. In this work, in order to produce different values of binding
energies, we focus on a particular way of varying the depth of the real 
$K^-$-nucleus potential by scaling down successively $g_{\sigma K}$ from 
its initial value $g^{(1)}_{\sigma K}$ and, once it reaches zero, scaling
down $g_{\omega K}$ too from its initial value $g^{(1)}_{\omega K}$
until the $K^-$ $1s$ state became unbound. Obviously, the good {\it global} 
fit to the atomic data is inevitably lost once the coupling constants are 
varied in order to scan over a wide range of binding energies. 
The reverse procedure of first scaling down $g_{\omega K}$ from its 
initial value $g^{(1)}_{\omega K}$ to zero, and then reducing 
$g_{\sigma K}$ from its initial value $g^{(1)}_{\sigma K}$ 
until the $K^-$ $1s$ state became unbound, gave very similar results except 
for weak binding where the calculated widths are larger than 100 MeV 
and hence are outside of our direct interest. 
Furthermore, in order to scan the region of large values of $B_{K^-}$, 
of order 200 MeV, we also scaled up $g_{\sigma K}$ from its initial value 
$g^{(1)}_{\sigma K}$ while keeping $g_{\omega K}=g^{(1)}_{\omega K}$. 
In Ref. \cite{MFG05} we have shown that similar results are obtained 
for different starting values for $g_{\sigma K}$ and $g_{\omega K}$, adjusted 
to a `shallow' variety potential \cite{ROs00}. We have also 
shown there the little sensitivity to the nuclear RMF model version chosen, 
the linear version L-HS \cite{HSe81} and the nonlinear version NL-SH 
\cite{SNR93}. Below we discuss some aspects of the sensitivity to the 
nonlinear version used, particularly for high nuclear densities. 

\begin{figure}
\includegraphics[height=9cm,width=10cm]{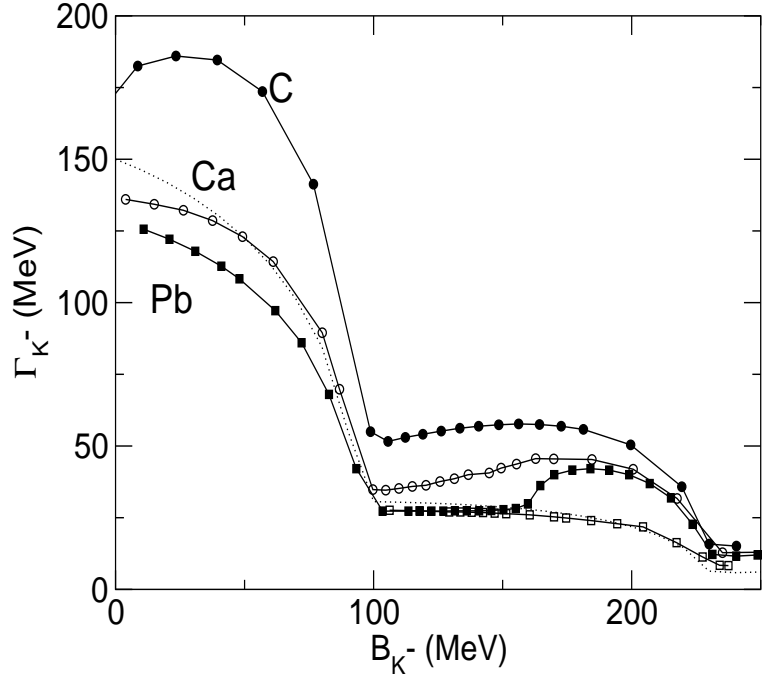}
\caption{Dynamically calculated widths of the $1s$ $K^-$-nuclear state
in $^{~~12}_{K^-}$C, $^{~~40}_{K^-}$Ca and $^{~~208}_{K^-}$Pb as function
of the $K^-$ binding energy, for the NL-SH RMF model {\protect \cite{SNR93}}. 
For $^{~~208}_{K^-}$Pb, results using the L-HS RMF model 
{\protect \cite{HSe81}} are also shown, by open squares. 
The dotted line is for a static nuclear-matter calculation with 
$\rho_0=0.16~{\rm fm}^{-3}$.}
\label{fig:Gamma}
\end{figure}

Figure \ref{fig:Gamma} shows calculated
widths $\Gamma_{K^-}$ as function of the binding energy $B_{K^-}$ for $1s$
states in $^{~~12}_{K^-}$C, $^{~~40}_{K^-}$Ca, $^{~~208}_{K^-}$Pb, for the
nonlinear NL-SH version \cite{SNR93} of the RMF model.
The dotted line shows the static `nuclear-matter' limit
\begin{equation}
\label{eq:Gamma}
\Gamma_{K^-}~=~\frac{f}{1-\frac{B_{K^-}}{m_K}}~\Gamma_{K^-}^{(0)}~~,
\end{equation}
where $f$ is the phase-space suppression factor of Eq. (\ref{eq:spf})
and $\Gamma_{K^-}^{(0)}$ is given by
\begin{equation}
\label{eq:rho0}
\Gamma_{K^-}^{(0)}~=~{\frac{4\pi}{\mu_{KN}}}~{\rm Im}b_0~\rho_0~~,
\end{equation}
for the static value ${\rm Im}b_0=0.62$ fm used in the calculations and
for $\rho_0=0.16$ fm$^{-3}$. Eq. (\ref{eq:rho0}) holds for a $K^-$ 
Schroedinger wavefunction which is completely localized
within the nuclear central-density $\rho_0$ region. The additional factor
$(1-B_{K^-}/m_K)^{-1}$ in Eq. (\ref{eq:Gamma}) follows from using the KG
equation rather than the Schroedinger equation. A small correction term
of order $V_{\rm c}/m_K$ was neglected in this factor.
It is clearly seen that the dependence of the width of the $K^-$ nuclear
state on its binding energy follows the shape of the dotted line
for the static nuclear-matter limit of $\Gamma_{K^-}$, Eq. (\ref{eq:Gamma}).
This dependence is due primarily to the binding-energy
dependence of the suppression factor $f$ which falls off rapidly until
$B_{K^-} \sim 100$ MeV, where the dominant
$\bar K N \rightarrow \pi \Sigma$ gets switched off, and then stays
rather flat in the range $B_{K^-} \sim 100 - 200$ MeV where the width is 
dominated by the two-nucleon absorption modes Eq. (\ref{eq:conv2}). 
The larger values of width for the lighter nuclei are due to the 
dynamical nature of the RMF calculation, whereby the nuclear density is 
increased by the polarization effect of the $K^-$ as shown in the next 
figures. We note that the widths
calculated in the range $B_{K^-} \sim 100 - 200$ MeV assume values
of about $50~-~60$~MeV for $^{~~12}_{K^-}$C, $35~-~45$~MeV for 
$^{~~40}_{K^-}$Ca, and about $25~-~30$~MeV for $^{~~208}_{K^-}$Pb in the 
range $B_{K^-} \sim 100 - 160$ MeV, decreasing gradually with $A$ to the 
`nuclear matter' dotted-line value of about 25 MeV. However, as is clearly 
observed in the figure, for $^{~~208}_{K^-}$Pb beginning at $B_{K^-} 
\sim 160$ MeV, a strong sensitivity to the nonlinear version of the RMF 
calculation develops whereby, in the NL-SH version \cite{SNR93} used here, 
the $^{208}$Pb s.p. levels undergo significant crossings, the maximal nuclear 
density increases substantially, and as  a result the nuclear rearrangement 
energy becomes significantly larger. This does not necessarily affect other 
bulk properties, such as the average nuclear density discussed below, 
that do not show a similar sensitivity. 
For this reason we also demonstrate in Fig. \ref{fig:Gamma} by open squares 
the results of using the linear version L-HS \cite{HSe81} of the RMF 
calculation for $^{~~208}_{K^-}$Pb above 100 MeV. The nonlinear and linear 
versions give practically the same results below 160 MeV, and we note 
assuredly that the behavior of the calculated width as function of the 
binding energy for the linear version follows closely the `nuclear matter' 
dotted line without any interruption around 160 MeV. 
The compressibility of the nucleus for linear versions is about twice that 
produced by nonlinear versions which are more realistic in this respect, 
and hence it is more difficult to polarize the nucleus within the linear 
version. 

\begin{figure} 
\includegraphics[height=9cm,width=10cm]{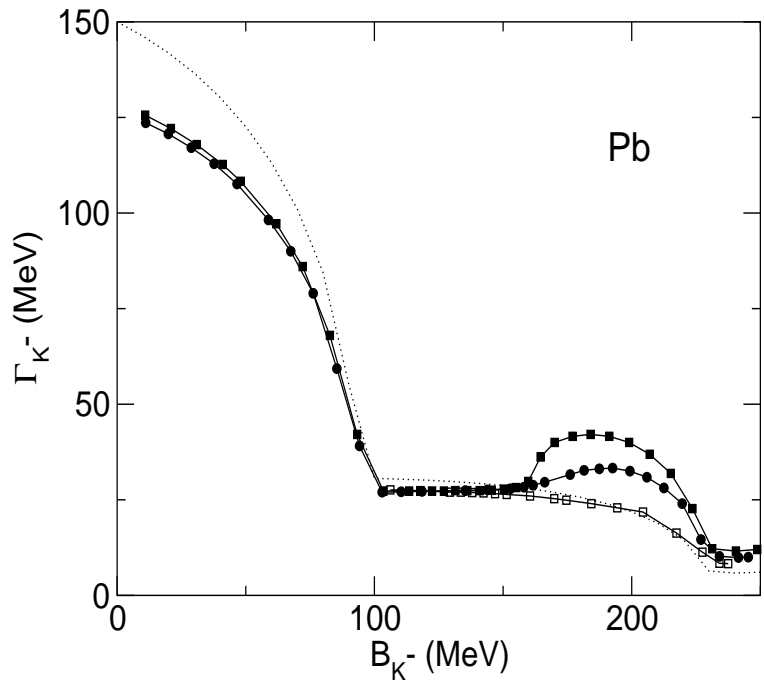} 
\caption{Dynamically calculated widths of the $1s$ $K^-$-nuclear state 
in $^{~~208}_{K^-}$Pb as function of the $K^-$ binding energy, for the 
nonlinear RMF models NL-SH {\protect \cite{SNR93}} (solid squares) and 
NL-TM1 {\protect \cite{STo94}} (solid circles), and for the linear model 
L-HS {\protect \cite{HSe81}} (open squares). The dotted line is for a static 
nuclear-matter calculation with $\rho_0=0.16~{\rm fm}^{-3}$.} 
\label{fig:Pb} 
\end{figure} 

In order to explore further the sensitivity to the nonlinear 
version of the RMF calculation, we show in Fig. \ref{fig:Pb} the calculated 
widths for $^{~~208}_{K^-}$Pb using in addition to the NL-SH version employed 
above \cite{SNR93} also the NL-TM1 version due to Sugahara and Toki 
\cite{STo94}. The results of using the linear L-HS version \cite{HSe81} are 
also shown for $B_{K^-} > 100$ MeV, as well as the dotted line for the static 
muclear-matter calculation specified above. The two nonlinear versions agree 
with each other very closely below 160 MeV but depart from each other above, 
with each giving rise to enhanced values of the width. The width evaluated 
using the NL-TM1 version appears to behave considerably more regularly than 
the NL-SH version . We note that similar calculations using the NL1 version, 
which was used in Ref. \cite{FGM99}, completely break down for the high 
nuclear densities encountered in this kinematical region. In contrast, the 
linear version L-HS \cite{HSe81}, with a considerably higher value of nuclear 
compressibility, does not give rise to any enhancement of the calculated width 
and it practically coincides with the results of the nuclear-matter static 
calculation. Another comment is that in the region below 100 MeV, the static 
nuclear-matter calculation produces higher values of width than any of the 
RMF versions, since it assumes a perfect overlap of the $K^-$ wavefunction 
with the central-density region.  

To conclude the estimate of $K^-$ widths we comment that replacing $\rho$ 
by ${\rho}^2$ for the density dependence of the non-pionic decay modes 
Eq.~(\ref{eq:conv2}) is estimated to increase the above values of the width 
by $10-15$ MeV. This estimate follows, again, from the increase of nuclear 
density with $B_{K^-}$ noticed above. 
Switching on the $\pi \Lambda$ decay mode would increase further this 
estimate by $5-10$ MeV in the range $B_{K^-} \sim 100 - 180$ MeV. 
We therefore expect that the estimate $\Gamma _{K^-} = 50 \pm 10$ MeV 
in the range $B_{K^-} \sim 100 - 200$ MeV provides a reasonable lower 
bound on the width expected for light and medium-weight nuclei in any 
realistic calculation.

\begin{figure}
\includegraphics[height=9cm,width=10cm]{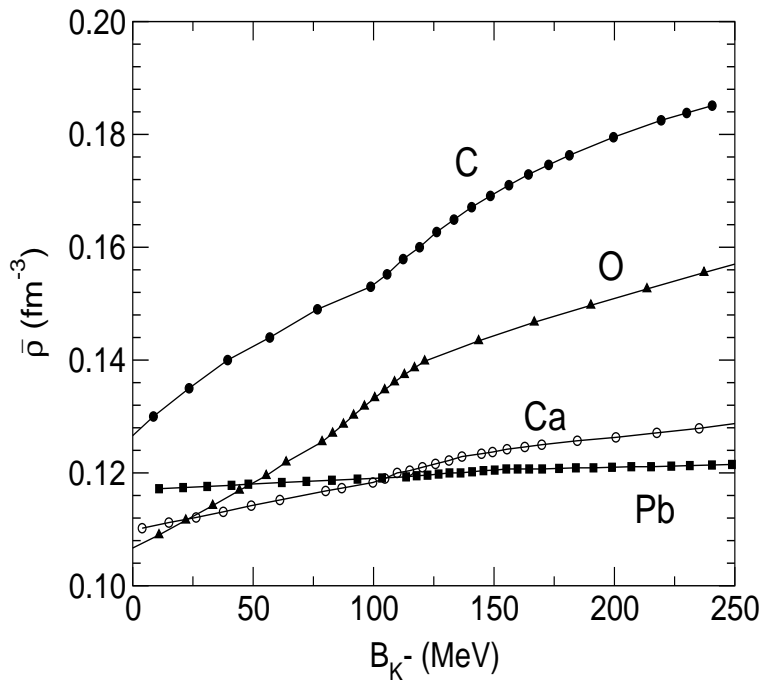}
\caption{Calculated average nuclear density $\bar \rho$ for
$^{~~12}_{K^-}$C, $^{~~16}_{K^-}$O, $^{~~40}_{K^-}$Ca and
$^{~~208}_{K^-}$Pb as function of the $1s$ $K^-$ binding energy, 
for the same nonlinear RMF model as in Fig. \ref{fig:Gamma}.}
\label{fig:rhobar}
\end{figure}

\begin{figure}
\includegraphics[height=9cm,width=10cm]{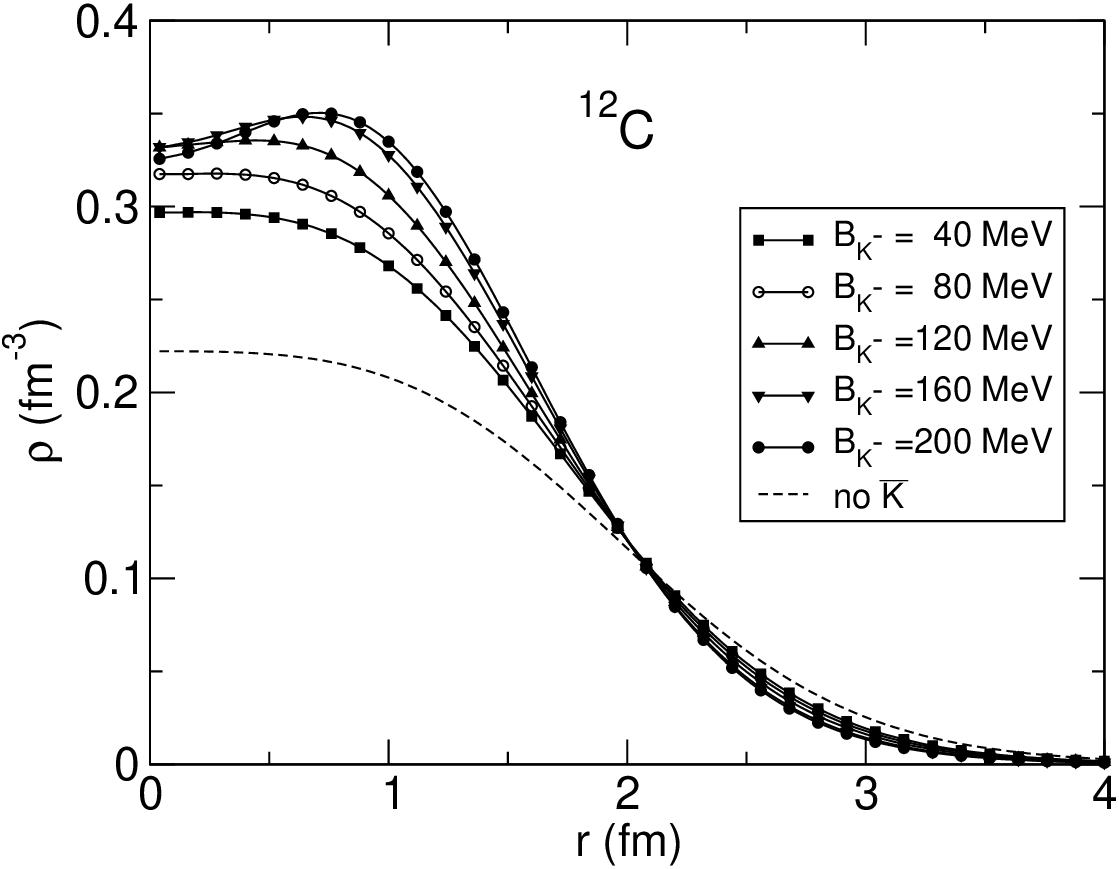} 
\caption{Calculated nuclear density $\rho$ of $^{~~12}_{K^-}$C 
for several $1s$ $K^-$ nuclear states with specified $B_{K^-}$ values, 
using the nonlinear RMF model NL-SH {\protect \cite{SNR93}} as in 
Fig. \ref{fig:Gamma}. The dashed curve stands for the $^{12}$C 
density in the absence of the $K^-$ meson.} 
\label{fig:Crho} 
\end{figure} 

\begin{figure}
\includegraphics[height=9cm,width=10cm]{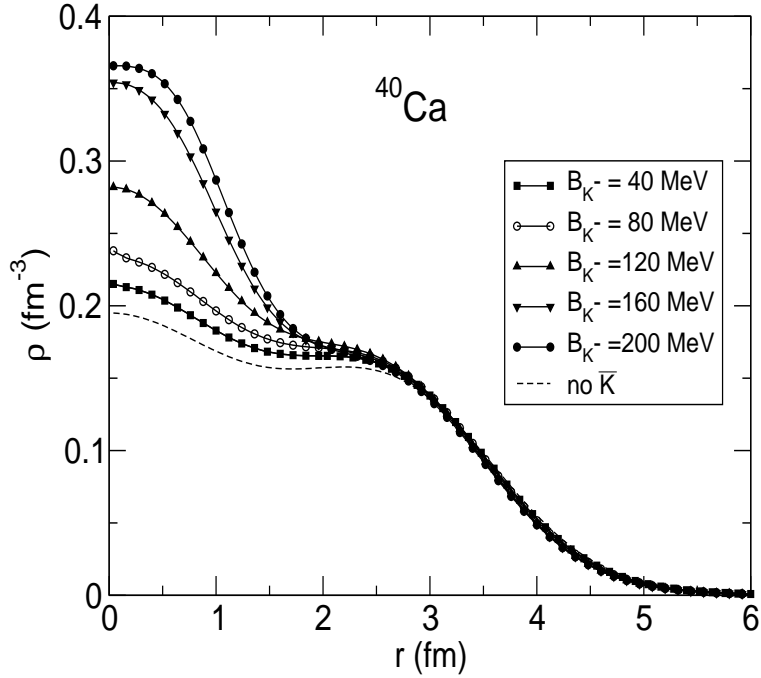}
\caption{Calculated nuclear density $\rho$ of $^{~~40}_{K^-}$Ca 
for several $1s$ $K^-$ nuclear states with specified $B_{K^-}$ values, 
using the nonlinear RMF model NL-SH {\protect \cite{SNR93}} as in 
Fig. \ref{fig:Gamma}. The dashed curve stands for the $^{40}$Ca 
density in the absence of the $K^-$ meson.} 
\label{fig:Carho} 
\end{figure} 

\begin{figure}
\includegraphics[height=15cm,width=10cm]{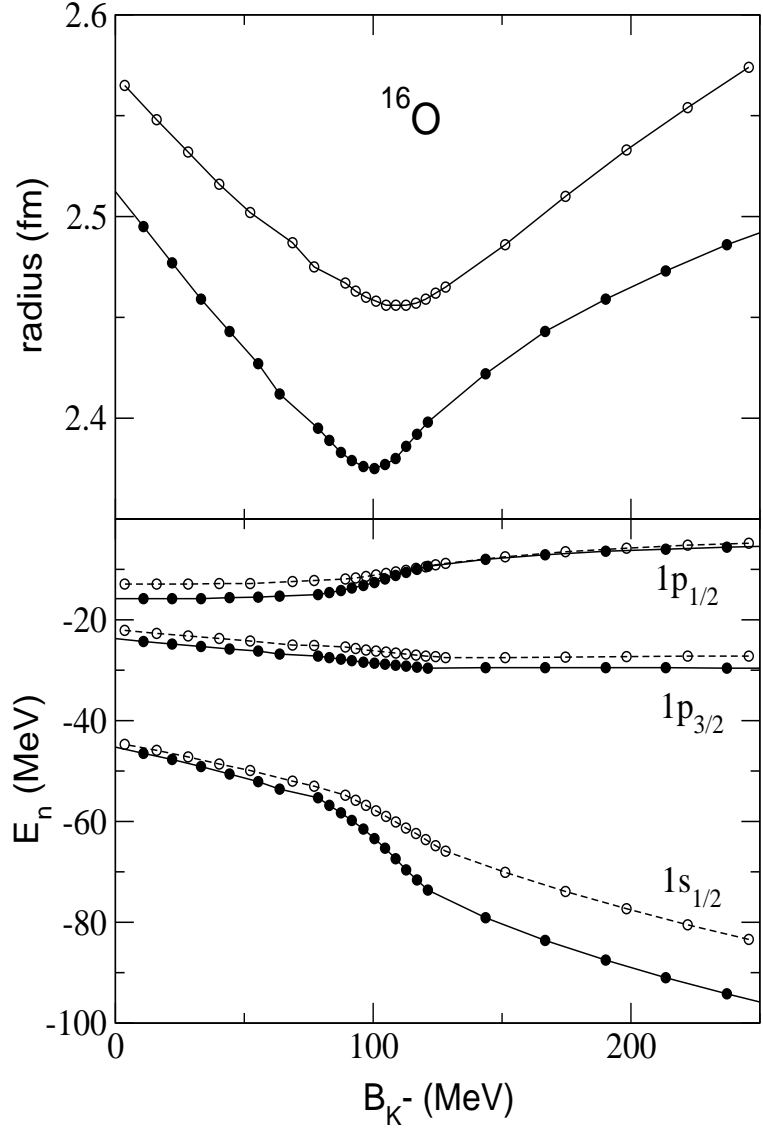}
\caption{Nuclear rms radius and neutron single-particle energies
for $^{~~16}_{K^-}$O as function of the $1s$ $K^-$ binding energy,
for the linear RMF model L-HS {\protect \cite{HSe81}} (open circles) and the
nonlinear RMF model NL-SH {\protect \cite{SNR93}} (solid circles) used in 
Fig. \ref{fig:Gamma}.}
\label{fig:nuclO}
\end{figure}

The next four figures exhibit various nuclear properties, calculated 
dynamically, for nuclei containing a $K^-$ nuclear $1s$ state. 
Figure \ref{fig:rhobar} shows the calculated average nuclear density 
$\bar \rho = \frac{1}{A}\int\rho^2d{\bf r}$ as a function of $B_{K^-}$ for 
the same $K^-$ nuclear $1s$ states as in Fig. \ref{fig:Gamma} and for $1s$ 
states in $^{~~16}_{K^-}$O which is studied below. 
The average nuclear density $\bar \rho$ increases substantially in the 
light $K^-$ nuclei, for the binding-energy range shown here, to values 
about 50\% higher than for these nuclei in the absence of the $K^-$ meson. 
The increase of the central nuclear densities is even bigger, 
as demonstrated in the next two figures, but is confined to a small region 
of order 1 fm from the origin. These features point out to a significant 
polarization of the nuclear core by the $1s$ $K^-$, particularly in light 
nuclei. 

Figure \ref{fig:Crho} shows calculated nuclear densities of 
$^{~~12}_{K^-}$C for several values of the $1s$ $K^-$ binding energies. 
The purely nuclear density, in absence of the $K^-$ meson, is given by the 
dashed curve. The maximal value of the nuclear density is increased by 
up to $75\%$ in the range of binding energies spanned in the figure, 
and the enhancement is close to uniform over the central 1 fm, decreasing 
gradually to zero by $r=2$~fm which already marks the nuclear surface. 
In this fairly small nucleus, the density is enhanced over a substantial 
portion of the nucleus. This is different than in heavier nuclei, 
as shown in Fig. \ref{fig:Carho} for the medium-weight $^{~~40}_{K^-}$Ca. 
Here, the enhancement of the maximal density (at the center of the nucleus) 
reaches almost a factor of two, but it subsides almost completely by 
$r=2$~fm which is still well within the nuclear volume (the nuclear radius 
of $^{40}$Ca is about 3.5 fm). This localization of the nuclear-density 
enhancement is due to the localized $1s$ $K^-$ density. As a result, 
the {\it average} nuclear density of $^{~~40}_{K^-}$Ca shown in 
Fig. \ref{fig:rhobar} is only weakly enhanced as function of $B_{K^-}$. 
In heavier nuclei such as $^{208}$Pb, the localization of the $1s$ $K^-$ 
density is as pronounced and the resultant nuclear density is enhanced 
by more than a factor of two over a small portion of the nucleus, confined 
again to the region $r \lesssim 2$~fm. 

The upper and lower panels of Fig. \ref{fig:nuclO} show the calculated
nuclear rms radius and the $1s$ and $1p$ neutron single-particle energies
$E_n$, respectively, for $^{~~16}_{K^-}$O as a function of $B_{K^-}$,
for two versions of the RMF calculation, the linear model L-HS \cite{HSe81} 
(open circles) and the nonlinear model NL-SH \cite{SNR93} (solid circles). 
The figure shows clearly that the polarization effect of the $1s$ $K^-$ 
bound state on the $1s$ nuclear bound state is particularly strong. 
The differences between 
the linear and nonlinear models reflect the different nuclear
compressibility and the somewhat different nuclear sizes obtained in the
two models. It is interesting to note that the increase in the
nuclear rms radius of $^{~~16}_{K^-}$O for large values of $B_{K^-}$ is
the result of the reduced binding energy of the $1p_{1/2}$ state,
due to the increased spin-orbit term. Note also that as $B_{K^-}$
approaches zero we do {\it not} recover the values inherent in static
calculations for the various nuclear entities in Fig. \ref{fig:rhobar}
and Fig. \ref{fig:nuclO}, because at $B_{K^-}=0$ both
Re$V_{\rm opt}^{\bar K}$ and Im$V_{\rm opt}^{\bar K}$ are still far
from assuming zero values. 

The polarization effects noted here in both Fig. \ref{fig:rhobar} and 
Fig. \ref{fig:nuclO} are somewhat larger than in Fig. 2 and Fig. 3 of 
our Letter publication \cite{MFG05}. This is due to the energy dependent 
suppressive factor in front of the vector potential in Eq.~(\ref{eq:VOP1}) 
which was disregarded there and which requires stronger couplings of the 
$\bar K$ in order to reproduce the same value of $B_{K^-}$ as before. 
However, the effect of this suppressive factor on the calculated widths is 
minor as long as the width is plotted as function of the binding energy, 
as done here in Fig. \ref{fig:Gamma} and in Fig. 1 of Ref. \cite{MFG05}, 
mainly because the dominant factor in the calculation of the width is the 
phase-space suppression factor $f$ which is directly determined by $B_{K^-}$. 
A welcome by-product of the present revision is that the `deep' potential 
fitted in section \ref{sec:katoms} to the $K^-$ atomic data produces $K^-$ 
nuclear bound states in light nuclei in the range of interest, 
$B_{K^-} \sim 100 - 200$~MeV, where several claims for bound states have 
recently been made \cite{SBF04,SBF05,KHA05,ABB05}. This is discussed 
in the next subsection. 

\subsection{Selected examples}
\label{example}

The discussion in this subsection is related to the
FINUDA experiment \cite{ABB05} at DA$\Phi$NE, Frascati, which recently
suggested evidence for a $K^-pp$ bound state produced in $K^-$
absorption at rest on unresolved combination of $^6$Li, $^7$Li and
$^{12}$C targets. The experiment looked for back-to-back $\Lambda p$
pairs, assigned to the strong decay $K^-pp \rightarrow \Lambda p$
which is a special case of Eq. (\ref{eq:conv2}). In addition to observing
$\Lambda p$ pairs which correspond kinematically to $K^-pp$ essentially
at rest, a peak in the invariant mass of $\Lambda p$ pairs was
found that corresponds in the interpretation of Ref. \cite{ABB05}
to a $K^-pp$ cluster bound by $B_{K^-pp} = 115 \pm 6 \pm 4$ MeV with
a decay width $\Gamma = 67 \pm 14 \pm 3$ MeV. We comment that the mere 
observation of back-to-back $\Lambda p$ pairs from the decay of a $K^-pp$ 
cluster does not mean that this cluster is self-bound in the nucleus, 
in as much as the observation of back-to-back $np$ pairs in the 
weak hypernuclear decay of $\Lambda$ hypernuclei does not mean that 
$\Lambda p$ clusters are self-bound in the nuclear medium. Since the 
$\bar K$-nucleus potential is at least as strong as the $\Lambda$-nucleus 
potential, the most natural expectation is for $\bar K$-nuclear bound states 
similar to the well accepted spectra of $\Lambda$-nuclear bound states 
over the periodic table beginning with $A=3$. The interpretation of the 
FINUDA events in terms of a $K^-pp$ bound state requires that its binding 
energy is independent of the nuclear target off which back-to-back 
$\Lambda p$ pairs are emitted. At present the data do not allow 
to extract a statistically significant signal from individual targets 
used in the experiment. Since the constitution 
of targets in the FINUDA experiment favors $^{12}$C, we will assume
for the sake of argument that this observed peak corresponds to
a $1s$ or a $1p$ $K^-$ bound state in $^{12}$C, or perhaps in $^{11}$C
or $^{11}$B corresponding to production by the $(K^-,n)$ or $(K^-,p)$
reactions respectively. The consequences of this working hypothesis
for the other targets will then be considered.
Subtracting $B_{pp} = 27.2$ MeV from the binding energy 115 MeV
assigned in Ref. \cite{ABB05} to the $K^-pp$-cluster binding,
the relevant $K^-$ binding energy is $B_{K^-} \sim 88$ MeV.
Regarding the width assigned by the FINUDA experiment to the 
bound-state signal, it could partly arise from Fermi-motion 
broadening which in the Fermi gas model yields the following contribution: 
\begin{equation}
\label{FGM} 
\frac{\Gamma_F}{2} = \sqrt{<{(\frac{p^2}{m_N})}^2> - {<\frac{p^2}{m_N}>}^2} 
= \frac{2}{5}{\sqrt\frac{3}{7}}\frac{p_F^2}{m_N} 
= 20.3~{\rm MeV} ~~~(p_F=270~{\rm MeV/c})~~.   
\end{equation}
This will reduce (quadratically) the assigned width of about 67 MeV 
down to 53 MeV. A rather broad distribution of back-to-back $\Lambda p$ 
events may also arise without invoking a bound state of any sort, due to 
the combination of Fermi motion of the initial $pp$ pair and the 
nuclear final-state interactions of the emitted $\Lambda p$ pair, as shown 
by Magas et al. \cite{MOR06} after submission of the present work for 
publication.  
    
\begin{table}
\caption{Binding energy $B_{K^-}$ calculated statically and
dynamically, nuclear rearrangement energy $R_{\rm nucl}^{\rm dyn}$
and width $\Gamma_{K^-}$ calculated dynamically,
for the $1s$ $K^-$ - nuclear bound state in $^{~~6}_{K^-}\rm Li$,
$^{~~12}_{K^-}\rm C$, $^{~~16}_{K^-}\rm O$, $^{~~40}_{K^-}\rm Ca$
and $^{~~208}_{K^-}\rm Pb$, using coupling-constant ratios
$\alpha_{\sigma} = g_{\sigma K}/g^{(1)}_{\sigma K} = 0$ and
$\alpha_{\omega} = g_{\omega K}/g^{(1)}_{\omega K} = 0.95$.}
\label{tab:1s}
\begin{ruledtabular}
\begin{tabular}{lcccc}
nucleus & $B_{K^-}^{\rm stat}$ (MeV)& $B_{K^-}^{\rm dyn}$ (MeV)&
$R_{\rm nucl}^{\rm dyn}$ (MeV)&$\Gamma_{K^-}^{\rm dyn}$ (MeV) \\
\hline
$^6$Li     & 28.1 & 35.6 & 17.0 & 229.1 \\
$^{12}$C   & 71.6 & 88.7 & 10.6 &  94.9 \\
$^{16}$O   & 66.0 & 72.6 &  7.6 & 100.4 \\
$^{40}$Ca  & 88.7 & 93.3 &  4.2 &  51.6 \\
$^{208}$Pb &107.3 &108.5 &  1.3 &  27.3 \\
\end{tabular}
\end{ruledtabular}
\end{table}

In Table \ref{tab:1s} we show results of static and dynamical calculations
for $1s$ states in $^{~~6}_{K^-}$Li, $^{~~12}_{K^-}$C, $^{~~16}_{K^-}$O,
$^{~~40}_{K^-}$Ca and $^{~~208}_{K^-}$Pb, choosing the coupling constants
$g_{\sigma K}$ and $g_{\omega K}$ such that $B_{K^-} \sim 88$ MeV for the
$1s$ ground state of $^{~~12}_{K^-}$C in the dynamical calculation. These 
coupling constants are not close to those required to produce the $K^-$ 
atoms best fit, and the $K^-$ potential determined here is considerably 
weaker than the potential of type DD or F in Fig.~\ref{fig:VR}.  
A comparison between the very different binding energies calculated for
$^{~~6}_{K^-}$Li and $^{~~12}_{K^-}$C leads to the conclusion that the
peak observed in the FINUDA experiment cannot result from {\it both} Li 
and C. The relatively weak binding calculated for $^{~~6}_{K^-}$Li implies,
in agreement with Fig. \ref{fig:Gamma} for other nuclei,
a huge value of over 200 MeV for the width. The width of the
$^{~~12}_{K^-}$C $1s$ state is smaller, $\Gamma = 95$ MeV, but is larger
than the width of the observed peak. These conclusions do not change upon
slightly increasing $g_{\omega K}$ from 0.95 to 0.985, so that
$B_{K^-} \sim 88$ MeV holds for $^{~~11}_{K^-}$C instead of for 
$^{~~12}_{K^-}$C. We then find $B_{K^-} = 88.6 (95.9)$ MeV and 
$\Gamma_{K^-} = 96.9 (66.5)$ MeV for $^{~~11(12)}_{K^-}$C, while 
$^{~~6}_{K^-}$Li is considerably less bound.

We note in Table \ref{tab:1s} the relatively strong variation of the
calculated binding energies between $^{~~12}_{K^-}$C and $^{~~16}_{K^-}$O
which is due to the difference between the underlying nuclear density input
for these nuclei in the static RMF calculation. Choosing only {\it one} of 
these nuclei for the mass-number $A$ systematics, the results of the table 
may be summarized as follows:
\begin{itemize}
\item
The $K^-$ binding energy increases monotonically as a function of $A$. 
However, without the Coulomb potential $V_c$ it is almost constant 
from $^{12}$C on due to saturation. 
\item
The width decreases monotonically with $A$, which is a corollary of the
$\Gamma (B_{K^-})$ dependence in Fig. \ref{fig:Gamma}.
\item
Except for $^{~~6}_{K^-}$Li where the binding energy is relatively small, 
the difference between the binding energies calculated dynamically and 
statically is substantial in light nuclei, decreasing monotonically with $A$, 
and may be neglected only for very heavy nuclei.
\item
The dynamical effect is very strong for $^{~~12}_{K^-}$C where the increase
in the dynamically calculated $B_{K^-}$ is 17 MeV with respect to the static
calculation. This increase gets larger with $B_{K^-}$, as demonstrated in the 
previous subsection for $\alpha_{\sigma} = \alpha_{\omega} =1$ where it 
amounts to 40 MeV.  
\item
The nuclear rearrangement energy, defined here as $R_{\rm nucl}^{\rm dyn}
= B_{K^-}^{\rm s.p.} - B_{K^-}$ in the dynamical calculation, decreases
monotonically with $A$. For a given value of $A$,
however (but not demonstrated here), $R_{\rm nucl}^{\rm dyn}$ increases
with $B_{K^-}$.
\end{itemize}

Except for $^{~~6}_{K^-}$Li, all the $K^-$ nuclei demonstrated in Table 
\ref{tab:1s} have additional bound states for the specified coupling constants. Thus, 
both $^{~~12}_{K^-}$C and $^{~~16}_{K^-}$O have bound $1p$ states, with 
$B_{K^-}^{\rm dyn} = 16.6,~30.6$ MeV, respectively, but with very large 
widths: $\Gamma_{K^-}^{\rm dyn} = 158,~123$ MeV, respectively. We therefore 
have an unfavorable situation of overlapping $1s$ and $1p$ levels.  

\begin{table}
\caption{Binding energy $B_{K^-}$ and width $\Gamma_{K^-}$ calculated
statically and dynamically for the $K^-$ - nuclear $1s$ and $1p$ states in
$^{~~6}_{K^-}\rm Li$ and $^{~~12}_{K^-}\rm C$, using
$\alpha_{\sigma} = 1.1$ and $\alpha_{\omega} = 1.0$.}
\label{tab:1s1p}
\begin{ruledtabular}
\begin{tabular}{lccccc}
nucleus & $nl$ & $B_{K^-}^{\rm stat}$ (MeV) & $\Gamma_{K^-}^{\rm stat}$ (MeV) 
& $B_{K^-}^{\rm dyn}$ (MeV) & $\Gamma_{K^-}^{\rm dyn}$ (MeV) \\
\hline
$^6$Li   & $1s$ & 75.1  & 79.0 & 146.5 &  74.5  \\
         & $1p$ &  4.1  & 70.7 &   7.4 & 184.4  \\
         &      &       &      &       &        \\
$^{12}$C & $1s$ &137.6  & 34.5 & 181.3 &  55.7  \\
         & $1p$ & 67.5  & 97.8 &  88.6 &  89.3  \\
\end{tabular}
\end{ruledtabular}
\end{table}

\begin{figure}
\includegraphics[height=12cm,width=12cm]{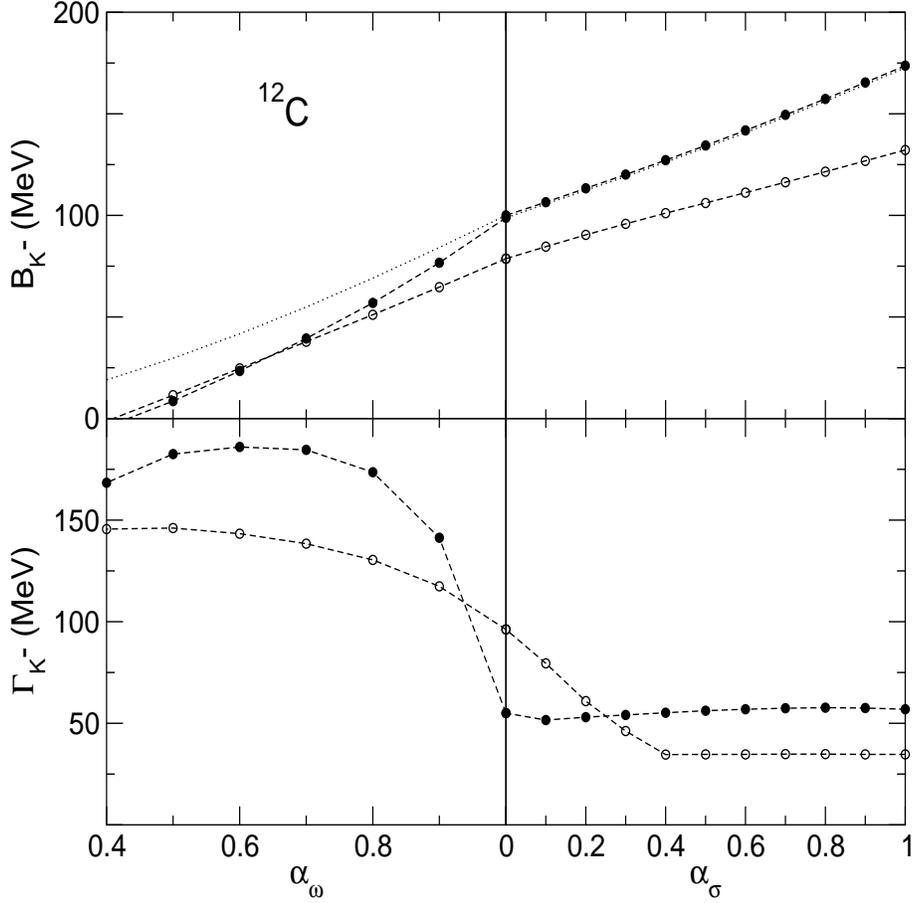}
\caption{$1s$ $K^-$ binding energy and width in $^{~~12}_{K^-}$C 
calculated statically (open circles) and dynamically (solid circles)
for the nonlinear RMF model NL-SH {\protect \cite{SNR93}} as function of
the $\omega K$ and $\sigma K$ coupling strengths: $\alpha_{\omega}$ is
varied in the left panels as indicated, with $\alpha_{\sigma}=0$,
and $\alpha_{\sigma}$ is varied in the right panels as indicated,
with $\alpha_{\omega}=1$. The dotted line shows the calculated binding
energy when the absorptive $K^-$ potential is switched
off in the dynamical calculation.}
\label{fig:dynam}
\end{figure}

In Table \ref{tab:1s1p}, we compare between dynamical calculations for
$^{~~6}_{K^-}$Li and $^{~~12}_{K^-}$C, requiring that $B_{K^-} \sim 88$ MeV
now holds for the $1p$ instead of the $1s$ state in $^{~~12}_{K^-}$C.
This requirement is motivated by the recent calculations of Yamagata et al.
\cite{YNO05} according to which the $1p$ production cross section in
$(K^-,N)$ reactions is higher than for the $1s$ state. The requirement 
$B_{K^-} \sim 88$ MeV for the $1p$ state in $^{~~12}_{K^-}$C is satisfied 
using kaon coupling constants only slightly different than those found 
optimal for the $K^-$ atomic fit of Ref. \cite{FGM99}. The width of this 
$1p$ state, 89 MeV, is close to the value 95 MeV listed in Table 
\ref{tab:1s} for the $1s$ state constrained to the same binding energy. 
The $1s$ states calculated in this case for the two nuclei 
are narrower than the respective $1p$ states, simply because their widths are 
suppressed stronger by the phase-space suppression factor $f$. The increase
in $B_{K^-}$ for these deeply bound $1s$ states provided by going from
the static calculation to the dynamical one is very substantial, 
with exceptionally large value of 71 MeV due to the strong polarization 
that the $^6$Li nucleus undergoes due to a deeply bound $K^-$. We note 
that for very shallow bound states, as demonstrated here for the 
$^{~~6}_{K^-}$Li $1p$ state and as will be shown in Fig. \ref{fig:dynam} 
below for the $^{~~12}_{K^-}$C $1s$ state, there is little difference between 
the static and the dynamical calculations for $B_{K^-}$. In contrast, the 
width $\Gamma_{K^-}$ is very strongly enhanced in $^{~~6}_{K^-}$Li owing to 
the increase of the nuclear density provided by the dynamical calculation, 
from 71 MeV in the static calculation to 184 MeV in the dynamical 
calculation for the $1p$ state. Although the width of the $1s$ state hardly 
changes, it would have decreased by roughly factor of two due to the 
phase-space suppression factor $f$ if the increase of the nuclear density 
did not offset this decrease.  

A comparison between 
the statically calculated (open circles) and the dynamically calculated
(solid circles) $B_{K^-}$ and $\Gamma_{K^-}$ for the $1s$ state in
$^{~~12}_{K^-}$C is shown in Fig. \ref{fig:dynam} as a function of the
coupling-constant strenghts. It is clear that for $B_{K^-} > 25$ MeV 
the dynamical calculation gives higher binding than the static calculation 
does, with the binding-energy gain increasing monotonically with $B_{K^-}$,  
and this effect becomes important for $B_{K^-} > 50$ MeV. The effect on 
$\Gamma_{K^-}$ is more intricate: the dynamically calculated width is 
larger than the width calculated statically, except in the region around 
$B_{K^-}~\sim~100$ MeV where the kinematical phase-space suppression 
factor $f$ of Eq. (\ref{eq:spf}) decreases rapidly. In this region, 
the dynamically calculated $\Gamma_{K^-}$ corresponds to binding energy 
$B_{K^-}$ which is associated with a considerably more suppressed value 
of $f$ than in the static calculation, and this effect wins over the width 
gained by the increased nuclear density. We note that the width calculated 
dynamically for the $1s$ nuclear state in $^{~~12}_{K^-}$C does not fall 
below 50 MeV for the range of variation shown, whereas the corresponding 
limiting value of the statically calculated width is about 35 MeV. Another 
feature shown in Fig. \ref{fig:dynam} concerns the effect of the imaginary 
potential on the binding energy: the dynamically calculated binding energy 
$B_{K^-}$ when Im$V_{\rm opt}$ is switched off is shown by the dotted line. 
It is clear that the absorptive potential Im$V_{\rm opt}$ acts 
{\it repulsively} and its inclusion leads to less binding, particularly 
at low binding energies. In fact, without Im$V_{\rm opt}$ we would 
{\it always} have $B_{K^-}^{\rm dyn} > B_{K^-}^{\rm stat}$. The repulsive 
effect of Im$V_{\rm opt}$ gets weaker with $B_{K^-}$, along with the action 
of the suppression factor $f$, and beginning with $B_{K^-}~\sim~100$~MeV it 
hardly matters for the calculation of $B_{K^-}$ whether or not 
Im$V_{\rm opt}$ is included.

Finally, we comment on the outcome of making $g_{\omega K}$ density 
dependent, 
\begin{equation}
\label{eq:WDD}
g_{\omega K}(\rho)~=~\frac{g_{\omega K}(0)}{1~-~0.046 \sigma \rho}~~, 
\end{equation} 
similar to the density dependence on the rhs in Eq.~(\ref{equ:WDD}),  
which according to Ref. \cite{Wei01} might simulate 
partial restoration of chiral symmetry in dense medium. 
The starting parameters for the static calculation corresponded to 
a `shallow' $\bar K$-nucleus potential of depths about 60 MeV for 
both real and imaginary parts. 
For this choice $^{~~12}_{K^-}$C and $^{~~16}_{K^-}$O have just one bound 
state ($1s$) each, with binding energy $B_{K^-}^{\rm dyn}=5.3,~22.8$ MeV, 
respectively. The introduction of the density dependence Eq.~(\ref{eq:WDD}) 
within a dynamical calculation increases the binding energy to about 
$70-80$ MeV. However, the width of these states is large, close to 100 MeV. 
We conclude that although this ansatz for density dependence did not improve 
the fit to kaonic atoms data where the need to introduce phenomenological 
density dependence appears at low densities, its role in the higher nuclear
density regime \cite{BRh96} and for generating nuclear $\bar K$
bound states cannot be ruled out at present.

\section{Summary}
\label{sec:sum}

In this work we re-analyzed the data on strong-interaction effects in $K^-$ 
atoms across the periodic table in order to study the density dependence of 
$V_{\rm opt}^{\bar K}$ and its extrapolation into nuclear-matter densities.
A departure from the $t_{\rm eff}\rho$ dependence was found at the $20\%$ 
range of nuclear-matter density $\rho_0$. Partial restoration of chiral 
symmetry, such as discussed in low energy pion-nuclear physics, was found 
to be ineffective at this low-density regime, but its role at higher 
densities cannot be ruled out. A smooth extrapolation into the nuclear 
interior gave a {\it deep} $\bar K$ nuclear potential $V_{\rm opt}^{\bar K}$ 
which would allow the existence of $\bar K$ nuclear bound states even for 
a static calculation. We then extended the purely nuclear RMF model by 
coupling the $\bar K$ {\it dynamically} to the nucleus. Negligible 
polarization effects were found for {\it atomic} states, which confirms 
the optical-potential phenomenology of kaonic atoms in general, and as 
a valid starting point for the present study in particular. Binding energies 
and widths of $\bar K$ {\it nuclear} bound states were then calculated across 
the periodic table with the aim of placing lower limits on the widths 
expected for binding energies in the range of $100-200$ MeV. 
Substantial polarization of the nucleus was found in light nuclei 
such as $^{12}$C and $^{16}$O for deeply bound $\bar K$ nuclear states, 
resulting in locally increased nuclear densities up to about $2\rho_0$. 
An almost universal shape was found for the dependence of the $\bar K$ 
width on its binding energy. 
The widths are primarily suppressed by the reduced phase-space for 
absorption of $\bar K$, and are enhanced by the increased density of the
polarized nucleus. The present results already provide useful guidance for 
the interpretation of the recent FINUDA experimental results \cite{ABB05} 
by presenting some calculated dependence on the atomic mass number $A$
and placing a lower limit $\Gamma_{\bar K} \sim 50 \pm 10$ MeV on $\bar K$ 
states bound in the range $B_{\bar K} \sim 100 - 200$ MeV. Our calculated 
widths for a $K^-$ $1s$ or $1p$ bound state in $^{12}$C, in the kinematical 
region of the FINUDA signal, are about 90 MeV which is somewhat larger than 
the reported width. We noted that a $1p$ bound state 
in $^{12}$C, rather than $1s$ bound state, is compatible with the 
`deep' type of $K^-$-nucleus potential fitted to the atomic data and 
perhaps also with deeply bound $K^-$ $1s$ states in the range 
$B_K \sim 150 - 200$ MeV for lighter nuclear targets such as $^4$He 
\cite{SBF04,SBF05}. We have argued that if the FINUDA observed signal 
corresponds to a $\bar K$ bound state in $^{12}$C, then it is unlikely 
to arise in Li targets, thus emphasizing the need to get statistically 
significant data from separate nuclear targets. 
For lighter nuclear targets, where the RMF approach becomes unreliable 
but where nuclear polarization effects are found much larger using
few-body calculational methods \cite{ADY05}, we anticipate larger
widths than 50 MeV for $\bar K$ deeply bound states in the range
$B_K \sim 100 - 200$ MeV.

\begin{acknowledgments}
This work was supported in part by the GA AVCR grant A100480617 
and by the Israel Science Foundation grant 757/05.
\end{acknowledgments}

\end{document}